\documentclass[12pt]{article}
\usepackage[left=2.5cm,top=2.50cm,right=2.5cm,bottom=2.50cm]{geometry}
\usepackage{mathrsfs}
\usepackage{amsmath,amssymb,latexsym,color,cancel,graphicx,bbm,colortbl}
\usepackage[english]{babel}
\usepackage[latin1]{inputenc}
\usepackage{ragged2e}
\usepackage{cite}
\begin{document}
\date{}

\title{Exact solutions of the Schr\"odinger Equation with Dunkl Derivative for the Free-Particle Spherical Waves, the Pseudo-Harmonic Oscillator and the Mie-type Potential}
\author{R. D. Mota$^{a}$ and  D. Ojeda-Guill\'en$^{b}$\footnote{{\it E-mail address:} dojedag@ipn.mx}} \maketitle

\begin{minipage}{0.9\textwidth}
\small $^{a}$ Escuela Superior de Ingenier{\'i}a Mec\'anica y El\'ectrica, Unidad Culhuac\'an,
Instituto Polit\'ecnico Nacional, Av. Santa Ana No. 1000, Col. San
Francisco Culhuac\'an, Del. Coyoac\'an, C.P. 04430, Ciudad de M\'exico, Mexico.\\

\small $^{b}$ Escuela Superior de C\'omputo, Instituto Polit\'ecnico Nacional,
Av. Juan de Dios B\'atiz esq. Av. Miguel Oth\'on de Mendiz\'abal, Col. Lindavista,
Del. Gustavo A. Madero, C.P. 07738, Ciudad de M\'exico, Mexico.\\

\end{minipage}

\begin{abstract}
We solve exactly the Schr\"odinger equation for the free-particle, the pseudo-harmonic oscillator and the Mie-type potential in three dimensions with the Dunkl derivative. The equations for the radial and angular parts are obtained by using spherical coordinates and separation of variables. The wave functions and the energy spectrum for these potentials are derived in an analytical way and it is shown that our results are adequately reduced to those previously reported when we remove the Dunkl derivative parameters.
\end{abstract}


\section{Introduction}

Dunkl introduced a combination of differential and difference operators, which are associated to a finite reflection group, and used them to study polynomials with discrete symmetry groups in several variables \cite{dunkl1,dunkl2}. The Dunkl derivative is closely related to the so-called Bannai-Ito and Dunkl-Schwinger algebras. Reflection operators were introduced by Wigner \cite{wigner} and applied to the one-dimensional harmonic oscillator by Yang \cite{yang}. These reflection operators have been used to study some properties of the quantum Calogero and Calogero-Sutherland-Moser models \cite{br,ply,hikami,kakei,lapon}.

The Dunkl derivative has been used to study some physical problems as the two-dimensional harmonic oscillator \cite{GEN1,GEN2} and the $2D$ Coulomb potential \cite{GEN3}. In the solution of these problems the Hermite, Laguerre and Jacobi polynomials have appeared. Moreover, in Refs. \cite{nos1,nos2}, it has been introduced the $su(1,1)$ Lie algebra and its irreducible representation to obtain the radial solutions and their coherent states. Also, the Dunkl-Coulomb and the harmonic oscillator problems for the Schr\"odinger equation in $3D$ have been solved and its superintegrability and dynamical symmetry have been studied \cite{GEN4,sami}.

The Dunkl derivative has also been used to study physical problems in the relativistic regime. In Refs. \cite{nos3,nos4}, the Dirac-Dunkl oscillators in one and two dimensions have been completely solved. Similarly, for the Klein-Gordon equation it is shown that the Coulomb potential, the Klein-Gordon oscillator and the Landau levels for the Klein-Gordon oscillator are exactly solvable in Refs. \cite{nos5,nos6}.

In the present work we obtain the exact solutions of the Schr\"odinger equation with Dunkl derivative for the free particle spherical waves, the pseudo-harmonic oscillator and the Mie-type potential.

This work is organized as follows. In Section $2$ the Hamiltonian for any general central potential is separated in its radial and angular parts in spherical coordinates. Then, we give the solution of the angular part in terms of the Jacobi polynomials and write a simplified version of the radial part. Section $3$ is dedicated to obtain the analytical solution of the free-particle spherical waves in terms the ordinary Bessel functions of half-integer order. The eigenfunction and energy spectrum for the pseudo-harmonic oscillator are obtained in Section $4$. The Mie-type potential is completely solved in Section $5$ and it is shown that our results are adequately reduced to the standard problem if we remove the Dunkl derivative parameters. Finally, we give some concluding remarks.

\section{Angular solutions of Schr\"odinger Equation with Dunkl Derivative }

In this Section we will obtain the general solution of the Schr\"odinger equation with the Dunkl derivative for any general central potential. These results are already known and have been reported in Ref. \cite{GEN4}. Hence, in what follows we set the mass of the particle $m=1$, $\hbar=1$, and $\omega=1$. Hence, for stationary states, the Schr\"odinger equation of a particle in a spherically symmetric potential $V(r)$ is
\begin{equation}
\left(-\frac{1}{2}{\nabla}^2+V(r)\right)\Psi(\vec r)={\mathcal E}\Psi(\vec r).
\end{equation}
When  we change the standard partial derivatives $\frac{\partial}{\partial x_i}$ by the Dunkl derivatives $D_i$
\begin{equation}
D_i\equiv\frac{\partial}{\partial x_i}+\frac{\mu_i}{x_i}(1-R_i), \hspace{5ex}i=1, 2, 3,
\end{equation}
we obtain the Schr\"odinger equation
 \begin{equation}
{\mathcal H}\psi(\vec r)\equiv\left(-\frac{1}{2}{\bf D}^2+V(r)\right)\Psi(\vec r)={\mathcal E}\Psi(\vec r) ,
\end{equation}
where we have defined the Hamiltonian $\mathcal H$, and where ${\bf D}^2=D_1^2+D_2^2+D_3^2\equiv \nabla^2_D$ is known as the Dunkl Laplacian.
The constants $\mu_1$, $\mu_2$ and $\mu_3$ satisfy $\mu_i>0$ \cite{GEN4}, and $R_1$, $R_2$  and $R_3$ are the reflection operators with respect to the $x_1-$, $x_2-$ and $x_3-$coordinates:
\begin{eqnarray}
&&R_1f(x_1,x_2,x_3)=f(-x_1,x_2,x_3),\\
&&R_2f(x_1,x_2,x_3)=f(x_1,-x_2,x_3),\\
&&R_3f(x_1,x_2,x_3)=f(x_1,x_2,-x_3).
\end{eqnarray}
Using spherical coordinates,
\begin{equation}
x_1=r\cos{\phi}\sin{\theta},\hspace{7ex}x_2=r\sin{\phi}\sin{\theta}\hspace{7ex}x_3=r\cos{\phi},
\end{equation}
the Hamiltonian $\mathcal H$ for any central potential $V(r)$ results to be
\begin{equation}
{\mathcal H}={\mathcal M}_r+\frac{1}{r^2}{\mathcal N}_\theta+\frac{1}{r^2\sin^2{\theta}}{\mathcal B}_\phi,
\end{equation}
where
\begin{eqnarray}
&&{\mathcal M}_r=-\frac{1}{2}\frac{\partial^2}{\partial r^2}-\frac{1+\mu_1+\mu_2+\mu_3}{r}\frac{\partial}{\partial r}+V(r),\label{mr}\\
&&{\mathcal N}_\theta=-\frac{1}{2}\frac{\partial^2}{\partial \theta^2}+\left(\mu_3 \tan{\theta}-\left(\frac{1}{2}+\mu_1+\mu_2 \right)\cot{\theta}\right)\frac{\partial}{\partial \theta}+\frac{\mu_3}{2\cos^2{\theta}}(1-R_3),\\
&&B_\phi\equiv-\frac{1}{2}\frac{\partial^2}{\partial \phi^2}+\left(\mu_1\tan{\phi}-\mu_2\cot{\phi}\right)\frac{\partial}{\partial \phi}
+\frac{\mu_1}{2\cos^2{\phi}} (1-R_1)+\frac{\mu_2}{2\sin^2{\phi}} (1-R_2).
\end{eqnarray}
The action of the reflection operators on any function $f(r,\theta,\phi)$ of the spherical coordinates is as follows
\begin{equation}
R_1f(r,\theta,\phi)=f(r,\theta,\pi-\phi),\hspace{2ex}R_2f(r,\theta,\phi)=f(r,\theta,-\phi),\hspace{2ex}R_3f(r,\theta,\phi)=f(r,\pi-\theta,\phi).
\end{equation}
By setting the variables separation $\Psi(r,\theta,\phi)=R(r)\Theta(\theta)\Phi(\phi)$ for the wave function, the Schr\"odinger equation is equivalent to the set of differential equations
\begin{eqnarray}
&&\left({\mathcal B}_\phi-\frac{k^2}{2}\right)\Phi(\phi)=0,\label{bfi}\\
&&\left({\mathcal N}_\theta+\frac{k^2}{2\sin^2{\theta}}-\frac{q^2}{2}\right)\Theta(\theta)=0,\\
&&\left({\mathcal M}_r+\frac{q^2}{2r^2}-\mathcal{ E}\right)R(r)=0,\label{master}
\end{eqnarray}
where $\frac{k^2}{2}$ and $\frac{q^2}{2}$  are the separation constants.

Explicitly, the eigenfunctions $\Phi(\phi)$ are labeled in terms of the eigenvalues ($s_1,s_2$), where $s_i=\pm 1$ are the eigenvalues of the
reflection operators ($R_1,R_2$), and are written in terms of the Jacobi polynomials $P_m^{(\alpha,\beta)} (x)$ as
\begin{equation}
\Phi_m^{(s_1,s_2)}(\phi)=\eta_m\cos^{e_1}\phi\sin^{e_2}\phi \hspace{0.2cm} P_{m-e_1/2-e_2/2}^{\mu_2-1/2+e_2,\mu_1-1/2+e_1}(\cos{2\phi}).\label{angular}
\end{equation}
Here, $(e_1,e_2)$ are the indicator functions for the reflection operators eigenvalues, defined as
\begin{equation*}
  e_i=\left\lbrace
  \begin{array}{l}
      0,\hspace{1.0cm}\text{if $s_i=1$  },\\
      1,\hspace{1.0cm}\text{if $s_i=-1$ },\\
  \end{array}
  \right.
\end{equation*}
$i=1, 2.$  If $s_1s_2=-1$, then $m$ is a positive half-integer, and if $s_1s_2=1$, then $m$ is a non-negative integer. The factors $\eta_m$ are the normalization constants explicitly given by
\begin{align}
\eta_m=&\sqrt{\left(\frac{2m+\mu_1+\mu_2}{2}\right)\left(m-\frac{e_1+e_2}{2}\right)!}\nonumber\times\\
&\sqrt{\frac{\Gamma\left(m+\mu_1+\mu_2+\frac{e_1+e_2}{2}\right)}{\Gamma\left(m+\mu_1+\frac{1+e_1-e_2}{2}\right)\Gamma\left(m+\mu_2+\frac{1+e_2-e_1}{2}\right)}}.
\end{align}
From these results it can be shown that the eigenvalues of equation (\ref{bfi}) take the form
\begin{equation}\label{mcuad}
k^2=4m(m+\mu_1+\mu_2).
\end{equation}
From the orthogonality relation of the Jacobi polynomials, it can be deduced that the angular wavefunctions $\Phi_m^{(s_1,s_2)}(\phi)$ satisfy \cite{GEN1,GEN4}
\begin{equation}
\int_0^{2\pi}\Phi_m^{(s_1,s_2)}(\phi)\Phi_{m'}^{(s'_1,s'_2)}(\phi)|\cos{\phi}|^{2\mu_1}|\sin{\phi}|^{2\mu_2}d\phi=\delta_{m,m'}\delta_{s_1,s'_1}\delta_{s_2,s'_2}.
\end{equation}

The eigenfunctions $\Theta(\theta)$ are labeled by the eigenvalues of the reflection operator $R_3$, $s_3=\pm 1$, and are given by
\begin{equation}
\Theta_\ell^{(s_3)}(\theta)=\iota_\ell\cos^{e_3}{\theta}\sin^{2m}{\theta} P_{\ell-e_3/2}^{(2m+\mu_1+\mu_2,\mu_3+e_3-1/2)}(\cos{2\theta}),
\end{equation}
with the values of the separation constant given by
\begin{equation}
q^2=4(\ell+m)\left(\ell+m+\mu_1+\mu_2+\mu_3+\frac{1}{2}\right).\label{qu}
\end{equation}
If $s_3=1$, then $\ell$ is a non-negative integer, and if $s_3=-1$, then $\ell$ is a positive half-integer. The normalization constant $\iota_\ell$ is given by
\begin{equation}
\iota_\ell=\left(\frac{(2\ell+2m+\mu_1+\mu_2+\mu_3+1/2)\Gamma(\ell+2m+\mu_1+\mu_2+\mu_3+1/2+e_3/2)(\ell-e_3/2)!}{\Gamma(\ell+2m+\mu_1+\mu_2-e_3/2)\Gamma(\ell+\mu_3+1/2+e_3/2)}\right)^{\frac{1}{2}}.
\end{equation}

The Dunkl angular momentum operators (rotation generator) are defined as
\begin{equation}
J_1=-i(x_2D_3-x_3D_2),\hspace{5ex}J_2=-i(x_3D_1-x_1D_3),\hspace{5ex}J_3=-i(x_1D_2-x_2D_1),
\end{equation}
which, together with the Hamiltonian $\mathcal H$, satisfy the following commutation relations
\begin{equation}
\left[J_j,J_k\right]=i\epsilon_{jk\ell}J_\ell(1+2\mu_\ell R_\ell),\hspace{5ex}\left[J_i,{\mathcal H}\right]=0.
\end{equation}
In fact, it has been shown that $J_3$ and $\bf J^2$ are the symmetries responsible for the separation of variables in spherical coordinates. Moreover, it has been shown that the operators
\begin{eqnarray}
&&J_3^2=2{\mathcal B}_\phi+2\mu_1\mu_2(1-R_1R_2),\\
&&{\bf J}^2=2\left({\mathcal N}_\theta+\frac{1}{\sin^2{\theta}}{\mathcal B}_\phi\right)+2\mu_1\mu_2(1-R_1R_2)++2\mu_2\mu_3(1-R_2R_3)\\
&&\hspace{5ex}+2\mu_1\mu_3(1-R_1R_3)+\mu_1(1-R_1)+\mu_2(1-R_2)+\mu_3(1-R_3),
\end{eqnarray} are diagonal on the separated wave function in spherical coordinates. Their respective eigenvalues are reported in Ref. \cite{GEN3}.

In what follows, we will obtain the exact solution of equation (\ref{master}) for the free particle spherical waves, the pseudo-harmonic oscillator, and the Mie-type potentials. To this end, we first substitute Eqn. (\ref{mr}) into Eqn. (\ref{master}) to obtain
\begin{equation}
\left(-\frac{d^2}{dr^2}-\frac{2(1+\mu_1+\mu_2+\mu_3)}{r}\frac{d}{dr}+\frac{q^2}{r^2}+2V(r)\right)R(r)=2{\mathcal E}R(r). \label{rad}
\end{equation}
If we define
\begin{equation}
G(r)\equiv r^{a}R(r), \quad\quad a=1+\mu_1+\mu_2+\mu_3,\label{ge}
\end{equation}
the differential equation for $G(r)$ results to be
\begin{equation}
\left(-\frac{d^2}{dr^2}+\frac{a^2-a+q^2}{r^2}+2V(r)\right)G(r)=2{\mathcal E}G(r).\label{master2}
\end{equation}
Using Eqn. (\ref{qu}), a simple calculation shows that factor of the centrifugal term can be factorized as
\begin{equation}
a^2-a+q^2=(2\ell+2m+\mu_1+\mu_2+\mu_3)(2\ell+2m+\mu_1+\mu_2+\mu_3+1)\equiv s(s+1),
\end{equation}
where $s$ is defined by
\begin{equation}
s\equiv 2\ell+2m+\mu_1+\mu_2+\mu_3. \label{ese}
\end{equation}
This result allows us to write Eqn. (\ref{master2}) as
\begin{equation}
\left(-\frac{d^2}{dr^2}+\frac{s(s+1)}{r^2}+2V(r)\right)G(r)=2{\mathcal E}G(r). \label{mas}
\end{equation}
This version for the radial part of the Dunkl-Schr\"odinger equation will be useful in the study of the pseudo-harmonic oscillator.

\section{The Free-Particle Spherical Waves with Dunkl derivative}

For the moment we focus our attention in the analytical solutions of the free-particle spherical waves.
Setting $V(r)=0$ in Eqn. (\ref{master2}) and with the definitions $\tilde G(r)=r^{-\frac{1}{2}}G(r)$ and $\rho\equiv \sqrt{2\mathcal{E}}r$, we obtain
\begin{equation}
\left(\frac{d^2}{d\rho^2}+\frac{1}{\rho}\frac{d}{d\rho}+1-\frac{(a-\frac{1}{2})^2+q^2}{\rho^2}\right){\tilde G(\rho)}=0.
\end{equation}
A direct computation shows that
\begin{equation}
 \left(a-\frac{1}{2}\right)^2+q^2=\left(s+\frac{1}{2}\right)^2,
\end{equation}
with $s$ defined as in equation (\ref{ese}). Since $s$ is in general a real number, we identify this differential equation with the
 first kind Bessel functions of arbitrary order $s+\frac{1}{2}$ and variable $\rho$. Thus, we get
\begin{equation}
{\tilde G(r)}=J_{s+1/2}(\sqrt{2{\mathcal E}}r),
\end{equation}
and consequently, substituting this result into equation (\ref{ge}), we obtain that the analytical free-particle solutions are given by
\begin{equation}
R_{{\mathcal E}\;s}(r)=r^{-a+\frac{1}{2}}J_{s+1/2}(\sqrt{2{\mathcal E}}r)=r^{-a+1}\sqrt{\frac{2\sqrt{{2\mathcal E}}}{\pi}}j_s(\sqrt{2{\mathcal E}}r).
\end{equation}
Here, we have introduced the spherical Bessel functions defined in terms of the one-half Bessel functions  as
\begin{equation}
j_\ell(kr)=\sqrt{\frac{\pi}{2kr}}J_{\ell+\frac{1}{2}}(kr),
\end{equation}
which obey the orthogonality relation in the generalized sense
\begin{equation}
\int_0^\infty j_\ell(kr) j_{\ell'}(k'r)r^2dr=\frac{\pi}{2k^2}\delta(k-k')\delta_{\ell \;\ell'}.
\end{equation}
This equation allows us to obtain the generalized orthonormality relation for the free-particle spherical radial solutions
\begin{equation}
\int_0^\infty R_{{\mathcal E}\;s}(r)R_{{\mathcal E'}\;s'}(r)r^{2(1+\mu_1+\mu_2+\mu_3)}dr=\frac{1}{2\mathcal E}\delta({\mathcal E}-{\mathcal E'})\delta_{s \;s'}.
\end{equation}
Therefore, we have exactly solved the radial part of the Schr\"odinger equation with Dunkl derivative for the free-particle spherical waves.

\section{The Pseudo-Harmonic Oscillator with Dunkl derivative}

In this Section we shall study the pseudo-harmonic oscillator, given by the potential $V(r)=Ar^2+\frac{B}{r^2}+C$. By substituting this expression into Eqn. (\ref{mas}), and rearranging we get
\begin{equation}
\left(\frac{d^2}{dr^2}-\frac{s(s+1)+2B}{r^2}-2Ar^2+2({\mathcal E}-C)\right)G(r)=0.  \label{harmonic}
\end{equation}
With the new variable $x\equiv (2A)^{\frac{1}{4}}r$, this equation takes the form
\begin{equation}
\left(\frac{d^2}{dx^2}-\frac{s(s+1)+2B}{x^2}-x^2+\frac{2({\mathcal E}-C)}{\sqrt{2A}}\right)G(x)=0. \label{pseudo}
\end{equation}
This equation has the same form as the differential equation
 \begin{equation}
u''+\left(4n+2\alpha+2-x^2+\frac{\frac{1}{4}-\alpha^2}{x^2}\right)u=0,\label{lebedev}
\end{equation}
whose solutions are given in terms of the Laguerre polynomials as follows \cite{LEB}
\begin{equation}
u(x)=C_0e^{-\frac{x^2}{2}}x^{\alpha+\frac{1}{2}}L_n^\alpha(x^2),\hspace{5ex} n=0,1,2,...\label{ansol}
\end{equation}
being $C_0$ a normalization constant. The comparison between equations (\ref{pseudo}) and (\ref{lebedev}) leads to the algebraic equations.
\begin{equation}
\frac{1}{4}-\alpha^2=-(s(s+1)+2B),\hspace{8ex}\frac{2({\mathcal E}-C)}{\sqrt{2A}}=4n+2\alpha+2.\label{ecspseudo}
\end{equation}
From the first expression we obtain
\begin{equation}
\alpha=\sqrt{\left(s+\frac{1}{2}\right)^2+2B}.\label{alfa}
\end{equation}
Therefore, the eigenfunctions $G_{n\,s}(x)$ of equation (\ref{pseudo}) are given by
\begin{equation}
G_{n\,s}(x) =C_0e^{-\frac{x^2}{2}}x^{\alpha+\frac{1}{2}}L_n^\alpha(x^2)  \hspace{5ex} n=0,1,2,...
\end{equation}
The normalization constant $C_0$ can be computed by using the integral
\begin{equation}
\int_0^\infty e^{-x^2}x^{2\alpha+1}\left[L_n^\alpha(x^2)\right]^2dx=\frac{\Gamma(n+\alpha+1)}{2n!},
\end{equation}
from which we obtain
\begin{equation}
C_0=\sqrt{\frac{2n!}{\Gamma\left(n+\sqrt{(s+\frac{1}{2})^2+2B}+1\right)}}.
\end{equation}
Moreover, from this result and equation (\ref{ge}), we obtain that the radial eigenfunctions $R_{n\,s} (x)$ must be normalized according to
\begin{equation}
\int_0^\infty R_{n\,s}(x)R_{n'\,s'}(x)x^{2(1+\mu_1+\mu_2+\mu_3)}dx=\delta_{n\,s}\delta_{s\,s'}.
\end{equation}

On the other hand, the energy spectrum for the pseudo-harmonic oscillator can be obtained by substituting the expression (\ref{alfa}) for $\alpha$ into the second Eqn. of (\ref{ecspseudo})
\begin{equation}
{\mathcal E}=\sqrt{2A}\left(2n+1+\alpha\right)+C=\sqrt{2A}\left(2n+1+\sqrt{\left(s+\frac{1}{2}\right)^2+2B}\right)+C.
\end{equation}
The pseudo-harmonic oscillator can be reduced to the standard $3D$ Dunkl oscillator by setting $B=0=C$. Hence, using the definition of $s$ (Eqn. (\ref{ese})) and setting $B=0=C$ we obtain
\begin{equation}
{\mathcal E}=\left(2\ell +2m+\mu_1+\mu_2+\mu_3+2n+\frac{3}{2}\right)\sqrt{2A}.
\end{equation}
Now, if we notice that for a particle of unit mass $\omega=\sqrt{2A}$, this result shows that the spectrum of the pseudo-harmonic oscillator reduces to that of the Dunkl harmonic oscillator \cite{GEN4}. Also, we see that given any $s$ and $B$, the first of equations (\ref{ecspseudo}) defines $\alpha$, i.e.
\begin{equation}
s(s+1)+2B=\left(\alpha-\frac{1}{2}\right)\left(\alpha+\frac{1}{2}\right).
\end{equation}
Therefore, we have shown that the pseudo-harmonic oscillator for the Dunkl-Schr\"odinger equation is an exactly solvable problem.

\section{The Mie-Type Potential with Dunkl derivative}

The Mie-type potential is given by $V(r)=-\frac{{\mathcal A}}{r}+\frac{{\mathcal B}}{r^2}+{\mathcal C}$. By substituting this potential into the radial Eqn. (\ref{rad}), and rearranging we get
\begin{equation}
\left(\frac{d^2}{dr^2}+2a\frac{d}{dr}-\frac{q^2+2{{\mathcal B}}}{r^2}+2\frac{{{\mathcal A}}}{r}-2({{\mathcal C}}-{\mathcal E}) \right)R(r)=0.
\end{equation}
By  performing the change of variable $x\equiv \sqrt{8({\mathcal C}-{\mathcal E})}r$, this equation takes the form
\begin{equation}
\left(x\frac{d^2}{dx^2}+2a\frac{d}{dx}-\frac{q^2+2{\mathcal B}}{x}+\frac{{{\mathcal A}}}{\sqrt{2({\mathcal C}-{\mathcal E})}}-\frac{x}{4}\right)R(x)=0.  \label{radial}
\end{equation}
It is known that the differential equation
\begin{equation}
xu''+(\beta+1-2\nu)u'+\left(n+\frac{\beta+1}{2}+\frac{\nu(\nu-\beta)}{x}-\frac{x}{4}\right)u=0,\label{generaleq}
\end{equation}
has the following eigenfunctions \cite{LEB}
\begin{equation}
u(x)=Ce^{-\frac{x}{2}}x^\nu L_n^\beta(x),\hspace{5ex} n=0,1,2,...\label{colsol}
\end{equation}
where $C$ is a normalization constant. By comparison of Eqns. (\ref{radial}) and (\ref{colsol}), we identify $u(x)$ with $R(x)$, and the following set of equations must be satisfied
\begin{equation}
\beta-2\nu+1=2a,\hspace{7ex}
\nu(\nu-\beta)=-\left(q^2+2{\mathcal B}\right),\hspace{7ex}
n+\frac{\beta+1}{2}=\frac{{{\mathcal A}}}{\sqrt{2({\mathcal C}-{\mathcal E})}}.  \label{rest}
\end{equation}
From the first two equations we find
\begin{eqnarray}
&&\nu=\frac{1}{2}-a+\frac{1}{2}\sqrt{4a^2-4a+4q^2+8{\mathcal B}+1},\\ \label{lagc1}
&&\beta=\sqrt{4a^2-4a+4q^2+8{\mathcal B}+1}.\label{lagc2}
\end{eqnarray}
Since $a^2-a+q^2=s(s+1)$, these equations can be written as follows
\begin{eqnarray}
&&\nu=\frac{1}{2}-a+\sqrt{\left(s+\frac{1}{2}\right)^2+2{\mathcal B}},\\ \label{nu}
&&\beta=2\sqrt{\left(s+\frac{1}{2}\right)^2+2{\mathcal B}}.\label{beta}
\end{eqnarray}
Explicitly, from Eq. (\ref{colsol}) we obtain that the eigenfunctions $R_{n\,s}(x)$ of the Mie-type potential are
\begin{equation}
R_{n\,s}(x)=Ce^{-\frac{x}{2}}x^\nu L_n^\beta(x),\hspace{5ex} n=0,1,2,...
\end{equation}
By using Eq. (\ref{beta}) and the Laguerre integral property
\begin{equation}
\int_0^\infty e^{-x}x^{\beta+1}\left[L_n^\beta(x)\right]^2dx=\frac{(2n+\beta+1)\Gamma(n+\beta+1)}{n!},
\end{equation}
we find that the normalization constant $C$ is given by
\begin{equation}
C=\sqrt{\frac{n!}{(2n+2\sqrt{\left(s+\frac{1}{2}\right)^2+2{\mathcal B}}+1)\Gamma(n+2\sqrt{\left(s+\frac{1}{2}\right)^2+2{\mathcal B}}+1)}}.
\end{equation}
Moreover, from this normalization we can show that the radial eigenfunctions $R_{n\,s}(x)$ must satisfy the following generalized orthogonality relation
\begin{equation}
\int_0^\infty R_{n\,s}(x)R_{n'\,s'}(x)x^{2(1+\mu_1+\mu_2+\mu_3)}dx=\delta_{n\,s}\delta_{s\,s'}.
\end{equation}
From the last of equations (\ref{rest}), we obtain that the energy spectrum for the Dunkl-Mie-type potential explicitly is
\begin{equation}
{\mathcal E}=-\frac{{\mathcal A}}{2\left(n+\frac{1}{2}(\beta+1)\right)^2}+{\mathcal C}=-\frac{{\mathcal A}}{2\left(n+\frac{1}{2}(\sqrt{\left(s+\frac{1}{2}\right)^2+2{\mathcal B}}+1)\right)^2}+{\mathcal C}.
\end{equation}
We emphasize that when ${\mathcal A}=1$, ${\mathcal B}=0$ and ${\mathcal C}=0$, the energy spectrum reduces to
\begin{equation}
{\mathcal E}=-\frac{1}{2\left(n+s+1\right)^2}=-\frac{1}{2\left(n+ 2\ell+2m+\mu_1+\mu_2+\mu_3 +1\right)^2},
\end{equation}
which is in full agreement to that reported in Ref. \cite{sami} for the Dunkl-Coulomb problem in three dimensions.

\section{Concluding Remarks}
In the present paper we separated the Schr\"odinger equation in its radial and angular parts for any general central potential in spherical coordinates. Then, we studied the radial part of the Dunkl-Schr\"odinger equation for the free-particle spherical waves, the pseudo-harmonic oscillator, and the Mie-type potential. It was shown that these three problems are exactly solvable, and we obtained their respective energy spectrum and eigenfunctions in an analytical way. Furthermore, we show that, when the Dunkl derivative parameters are removed, our results are correctly reduced to those previously obtained for each of these problems.

\section*{Acknowledgments}
This work was partially supported by SNI-M\'exico, COFAA-IPN, EDI-IPN, EDD-IPN, and CGPI-IPN Project Number $20210734$.

\end{document}